\documentclass[manuscript]{aastex}

\shorttitle{Identifying BHB Stars Using the z Filter}
\shortauthors{Vickers et al.}

\begin{document}

\title{Identifying Blue Horizontal Branch Stars Using the z Filter}

\author{John J. Vickers\altaffilmark{1}, Eva K. Grebel, and Avon P. Huxor}
\affil{Astronomisches Rechen-Institut, Zentrum f\"{u}r Astronomie der Universit\"{a}t Heidelberg, M\"{o}nchhofstr. 12-14, 69120 Heidelberg, Germany}

\altaffiltext{1}{jvickers@ari.uni-heidelberg.de}

%
%
\begin{abstract}
In this paper we present a new method for selecting blue horizontal branch (BHB) candidates based on color-color photometry. We make use of the Sloan Digital Sky Survey $z$ band as a surface gravity indicator and show its value for selecting BHB stars from quasars, white dwarfs and main sequence A type stars. Using the $g$, $r$, $i$, and $z$ bands, we demonstrate that extraction accuracies on par with more traditional $u$, $g$, and $r$ photometric selection methods may be achieved. We also show that the completeness necessary to probe major Galactic structure may be maintained. Our new method allows us to efficiently select BHB stars from photometric sky surveys that do not include a $u$ band filter such as the Panoramic Survey Telescope and Rapid Response System.
\end{abstract}


%
%
\section{Introduction}Blue horizontal branch stars (BHBs) are helium core-burning stars, typically about 0.7M$_{\sun}$ and 3R$_{\sun}$, which exist at a near-constant absolute g band magnitude of about 0.7 \citep{yan00}. Since they are such luminous objects with a predictable brightness, they are frequently used as standard candles to explore distant Galactic structure and are widely sought tracers for structural and kinematic studies of the Milky Way (e.g. Clewley \& Jarvis 2006, Ruhland 2011, Wilhelm et al. 1999, Xue et al. 2008). Owing to their old ages ($\geq$12 Gyr in many globular clusters, Dotter et al. 2010), BHB stars are particularly excellent halo tracers, since the halo is expected to be an old, metal-poor structure. Other high luminosity stars, such as K and M giants, are less desirable as their absolute magnitudes cover a wide range of values and they may intrinsically be much younger objects.

Photometrically, BHBs reside in a specific color range, being just bluer in $g-r$ than the instability strip (RR Lyraes) and just redder than extended vertical blue horizontal branch (sub-dwarf) stars. This "blue" color slice (-0.3 $<$ $g-r$ $<$ 0.0; all magnitudes and colors in this paper are extinction corrected and dereddened according to the values of Schlegel et al. 1998 unless otherwise stated) contains three main contaminants to BHB selection: distant quasars, foreground white dwarfs, and main sequence A stars (MSA; blue stragglers are photometrically very similar to MSA stars and so they are included in the same category for the purposes of this paper).

The most accurate way to separate a BHB star from these contaminants is via spectroscopy. Spectral templates will immediately eliminate white dwarf and quasar contamination (as white dwarfs have very broad absorption lines, and quasars have strong emission rather than absorption lines). To separate MSA stars from BHBs, two main approaches exist: first a comparison of Balmer line depth ($f_{m}$) and broadening ($D_{0.2}$) caused by the different surface temperatures and surface gravities, respectively (see Pier 1983), of the stars; and secondly the scale-width-shape method described by \citet{cle02} which separates the two species based on the Balmer line fits to a S\'{e}rsic profile. 

In the absence of spectroscopic data however, photometric methods may be applied to separate BHB stars from other blue species. This has been shown to be quite effective in several studies. For example: \citet{yan00} demonstrated a filter cut in the Sloan Digital Sky Survey (SDSS; York et al. 2000) which separates BHB stars from MSA stars adequately enough to discern significant structure in the plane of the celestial equator; \citet{sir04} use a "stringent" cut similar to that proposed by \citet{yan00} on the basis of spectroscopic data and a combination of the scale-width-shape method and the $D_{0.2}$  and $f_{m}$ methods in their kinematic studies of the Galactic halo; more recently \citet{bel10} further refine this "stringent" cut in their investigation of the ratios of BHB to main sequence turn off stars in the halo.

For the SDSS, \citet{len98}, using Kurucz model spectra, suggest that optimal gravitational separation for blue A-type stars lies in the $u-g$ color space. This is largely due to the Balmer jump, a spectral feature at 365 nm which is highly dependent on surface gravity. \citet{len98} also suggest separation in $i-z$ space, but note that gravitational splitting in this color-space is less effective than in the $u-g$ color space. We show that splitting here is caused by the Paschen features which reside in the z band and are also sensitive to surface gravity.

This gravitational splitting in the $z$ band is a serendipitous development with regards to the ongoing Panoramic Survey Telescope and Rapid Response System (Pan-STARRS; Kaiser et al. 2010) project, which features filters similar to the SDSS photometric system ($ugriz$, see Fukugita et al. 1996, Gunn et al. 1998). The main difference being that the SDSS, originally designed as a primarily extragalactic survey, uses the $u$ band to efficiently separate out quasars (e.g. Richards et al 2009 and references therein) while Pan-STARRS is primarily concerned with near earth objects and, as such, opts for a near infrared $y$ band instead of the u filter. It is worth noting that Pan-STARRS, a purely photometric survey, plans to survey three-fourths of the celestial sphere to 24$^{th}$ magnitude in the optical \citep{kai10} while SDSS, as of Data Release 8 (DR8; Aihara et al. 2011) has imaged one-third of the sky to a limiting magnitude of 22.2 in g and r. So, by developing an efficient way to select BHB candidate stars in the absence of a u band, we may facilitate statistical structural studies in a larger area of sky, to fainter magnitudes, than is currently possible with SDSS data.

With this in mind, we create a color space cut to select out BHB stars from the primary contaminants based on surface gravity measurements, but rather than exploiting u band separation, we explore the usefulness of the z band. With an end-goal of structure mapping and halo studies in mind, we aim primarily for a high purity selection at the cost of completeness. We attempt to achieve accuracies similar to those quoted by \citet{sir04} and \citet{bel10} ($<$ 30\% contamination), however, we also expect a lowered sample completeness because of the poorer gravitational separation in this color space. In \S 2 we describe the formulation of the color cut using SDSS spectroscopic data. In \S 3 we estimate the completeness of the cut by examining a sample of 10 globular clusters located in the SDSS footprint. In \S 4 we explore the usefulness of this method for probing Galactic structure by examining photometric data of the celestial equator which is rife with features. In \S 5 we discuss the implications of this method.

%
%
\section{The Color Cut}
To investigate the separation of various blue objects in the $i-z$ color space suggested by \citet{len98}, we select spectra from SDSS DR8. DR8 extends the SDSS footprint to now cover a full third of the celestial sphere and increases the total number of spectra to over 1.8 million\footnote{http://www.sdss3.org/dr8/}. 

We selected spectroscopic data from the entire SDSS footprint, only clean (not near saturated pixels) objects evaluated as point sources were selected. We used only primary measurements (in the case of multiply observed objects, the highest signal to noise reading is flagged as the primary one). To make our study consistent with those of \citet{sir04} and \citet{bel10} we select only stars with $g$ $<$ 18.

In Figure 1, we plot the adopted log(g) value as calculated by the SEGUE Stellar Parameters Pipeline (SSPP, see Lee et al. 2008; SEGUE stands for the "Sloan Extension for Galactic Understanding and Exploration", see Yanny et al. 2009) against $g-r$. We note that in the blue color space of -0.3 $<$ $g-r$ $<$ 0.0 there is excellent differentiation of the lower gravity BHB stars, residing in the range of 3.0 $<$ log($g$) $<$ 3.75, and the higher gravity MSA stars (3.75 $<$ log($g$) $<$ 5.0). These color and gravity ranges form the basis for our identification of these two types of star.

Some contaminants have no SSPP atmospheric parameters, so we cannot identify them based on their log(g) estimates: the two principal examples are white dwarf stars and quasars. White dwarf stars raise critical flags in the SSPP because of the width of their Balmer lines ($D_{0.2}$  $>$ 35.0 \AA) and quasars raise critical flags due to their strong emission lines \citep{lee08}. To identify these contaminants, we rely on the ELODIE template matches (the ELODIE archive is a set of high resolution spectroscopic readings collected using the ELODIE spectrograph, which has been operating on the Observatoire de Haute-Provence 1.93 m telescope since 1993; Moultaka et al. 2004) as output by the SSPP. The remaining contaminants are binned together-- they mostly consist of A and F stars for which the SSPP gravity reading was either inconclusive or outside the bounds of the prior mentioned BHB and MSA star bins.

Using this data set we look to construct a photometric color cut in $g-r$ (a temperature indicator) vs $i-z$ (a surface gravity indicator) color space (see Figure 2). To do this we use a k-nearest-neighbors classification algorithm. This algorithm classifies unknown objects based on their proximity to known objects. We compare a uniform grid in $g-r$ vs $i-z$ color space (the grid consisting of 101x101 nodes over the color space -0.3 $<$ $g-r$ $<$ 0.0 and -0.25 $<$ $i-z$ $<$ 0.05) to the spectroscopic data-- grid nodes with 50\% of their 5 nearest neighbors having been classified as BHB stars were said to reside in BHB color space. We then drew a rough selection box around this BHB color space defined by the points: ($g-r$,$i-z$) = (-0.30,-0.18), (-0.05,-0.06), (-0.05,-0.02), (-0.30,-0.02).

In Figure 2 we plot these 5 datasets: high gravity MSA stars, low gravity BHB stars, spectroscopically identified white dwarfs, spectroscopically identified quasars and inconclusive points. The lower panel of the plot is a $g-r$ vs $i-z$ color-color plot: BHB stars are plotted with diamonds while contaminants are plotted with points. Note that only 1 in 5 data-points are shown in the lower panel of the plot to avoid obscuration. However, the upper histogram shows all of the objects which are inside the selection box. It is apparent that this separation culls white dwarfs with acceptable efficiency. Note that type DC white dwarfs were neglected from this analysis due to a lack of a SSPP "DC white dwarf" classification and their intrinsically low numbers ($<$4\% of all white dwarfs). Additionally, one could use a proper motion diagram to remove some of the remaining white dwarfs and some of the closer MSA stars.

We do not get much more separation from the stellar contaminants in other color spaces. However, we can more efficiently select out quasar contaminants in the $g-r$ vs $g-z$ color space. We use $g-z$ color space due to the spectral profiles of blue stars which are characterized by the tail end of blackbody profiles (tending to lower $g-z$ values) and quasars which are more uniform emitters (tending to higher $g-z$ values-- especially in the case of high redshifts). In Figure 3 we present this additional color cut. This cut is defined by the points: ($g-r$,$g-z$) = (-0.3,-0.72) , (-0.3,-0.57) , (-0.05,-0.08) , (-0.05,-0.23). We plot only the data which pass the cut shown in Figure 2. Quasars are plotted as open circles and stellar sources are plotted as dots to accentuate the separation. Again histogrammed above are the objects passing the color cut. As expected, we see very little change in the stellar contaminants, but quasar contamination drops significantly.

As a test of accuracy, we consider the number of stars falling within the color cut which are not BHB stars: out of $\sim$4300 spectra which pass this cut, we find $\sim$77\% of them to be BHB stars; this represents a sample of $\sim$74\% of the stars originally identified as BHB stars. For comparison, we run a similar test on the same data using the color cut suggested by \citet{bel10} which is similar to the "stringent"  color cut employed by \citet{sir04} ($0.98 <$ $u-g$ $<$ 1.28 , $-0.2 <$ $g-r$ $< -0.06$ , excluding $([u-g-0.98]/0.215)^2+([g-r+0.06]/0.17)^2 < 1$). Using this refined u band dependent test, we similarly select a sample that is $\sim$74\% pure and $\sim$72\% complete. This test is biased by the SDSS spectroscopic selection algorithms, and so is neither a strict test of purity nor completeness.

%
%
\section{The Globular Cluster Test}
We use a completeness test similar to the one employed by \citet{bel10}. Using \citet{jor10} as a reference, we select 10 globular clusters from the SDSS footprint with pronounced BHBs. By running the constituent stars through the photometric cut described above, we may get a second measure of how effective the algorithm is at extracting BHB stars. The clusters chosen and pertinent information is tabulated in Table 1.

Since the SDSS $photoObjAll$ pipeline fails for crowded fields, the cores of dense objects, such as globular clusters, are often omitted from the general photometric data. For accurate and complete cluster photometry we turn to the SDSS "value added" catalogs\footnote{http://www.sdss.org/DR7/products/value\_added/index.html}-- in particular we use the "$ugriz$ DAOPHOT photometry of SDSS+SEGUE Globular and Open Clusters" catalog produced by \citet{an08}.

Since the \citet{an08} value added catalog was created with SDSS Data Release Seven (DR7; Abazajian et al. 2009) imaging and pipelines, we use DR7 photometric data for this cluster analysis. Data are selected from DR7 within one third of the tidal radii of the relevant clusters (Harris (1996) Catalog, 2010 revision) and then matched to the DAOPHOT photometry. In the cases of duplicate points, DAOPHOT photometry is preferred. The stellar magnitudes and colors are extinction corrected and dereddened according to the Harris Catalog E(B-V) values.

By examining the color-magnitude diagrams of the clusters, we visually select 1-magnitude wide boxes encompassing the main portion of the blue horizontal branch in the -0.3 $<$ $g-r$ $<$ 0.0 regime. This selection box contains the stars we will consider "true" BHBs. All stars outside of these boxes will be considered "false." It is apparent from clusters such as NGC 7078 and NGC 5272 that these 1-magnitude boxes suffer RR Lyrae contamination on the red end and in the clusters NGC 6205 and NGC 6341 we see hot sub-dwarf contamination on the blue end. However, in general, these boxes should consist mainly of BHB stars. The ranges for these magnitude boxes are also given in Table 1 and the cluster color-magnitude diagrams with the selection boxes superimposed are shown in Figure 4.

We select only stars passing the color-cut selection described in \S 2. We do not consider this test to be a test of accuracy due to the selective enrichment of BHB stars in these cluster-fields-- however, we do consider this to be a good test of completeness. Stars passing our $z$ based color cut constitute a sample that is $\sim$95\% pure and $\sim$51\% complete. Similar to \S 2, we cut the data using the $u$ based "stringent" color selection and find a sample that is $\sim$92\% pure and $\sim$57\% complete (see Table 2).

We consider the statistics on the spectroscopic data to be a better indicator of accuracy and the globular clusters to be a better indicator of completeness. Thus, when considering this test in conjunction with the spectroscopic statistics, we consider the $u$ based cut to be $\sim$74\% pure and $\sim$57\% complete which is in agreement with the values quoted by \citet{bel10} and \citet{sir04}. The $z$ based cut similarly has a $\sim$77\% purity and a $\sim$51\% completeness. 

%
%
\section{The Celestial Equator}
To show the ability of this selection method to probe features of the Milky Way's structure, we examine the celestial equator. This area (-1.26$^{\circ}$ $<$ declination $<$ 1.26$^{\circ}$) has been almost completely imaged in the SDSS, excepting areas too close to the Galactic plane. The portion of this great circle above the Galactic equator is known as stripe 10 (see \citet{sto02} for stripe naming conventions in the SDSS) and the portion residing in the southern Galactic hemisphere is stripe 82. Stripe 82 has been imaged repeatedly to promote studies of variable objects, such as supernovae \citep{fri08}, RR Lyraes and quasars. As such, this section of the sky has been extensively probed for evidence of structure (see \citet{new02} and \citet{ses07} for examples).

We create our sample by selecting all photometrically clean sources identified as point sources (selecting only primary measurements in the case of duplicates) residing in the plane of the celestial equator and pass their dereddened photometric data through the color cut suggested in \S 2. We note that, in general, noise in spectra of stars fainter than $g$=18 throw errors into the parameter determinations (Sirko et al. 2004; exceptions include SEGUE pencil beams which were imaged for various time scales to maximize signal to noise ratios at two main magnitude bins, see Yanny et al. 2009)-- but since the CCD camera can reliably determine every color to much fainter magnitudes \citep{gun98}, photometric separation is still practical. Photometric separation has the parallel benefit of being viable for a much more complete sample of stars when compared to spectroscopic methods.

The results of our photometric selection are plotted in Figures 5 (the northern Galactic hemisphere portion of the celestial equator) and 6 (the southern Galactic hemisphere along the celestial equator). Immediately evident are several well documented structures: among these are the Sagittarius stream in both the northern and southern Galactic hemispheres \citep{yan00}, the Virgo overdensity \citep{viv01} in the north, and the Hercules-Aquila cloud in both the north and the south noted independently by \citet{new02} and \citet{bel07}, and the anomalous density at R.A. = 160$^{\circ}$ noted by \citet{new02}.

%
%
\section{Spectral Interpretation}
We suspect that the $z$ band shows gravitational separation due to the gravity sensitive Paschen features (the Paschen analog to the Balmer jump resides at 8201\AA), which lie in the $z$ band. We anticipate that this separation may be even more effective on the Pan-STARRS telescope which has a much more infrared sensitive CCD than the SDSS imaging camera, and thus a more responsive $z$ filter \citep{stu10}. Figure 7 depicts spectra of the objects discussed in this paper (BHB stars, MSA stars, white dwarfs, and quasars) and includes the throughput functions of both the SDSS 2.5 m telescope and the Pan-STARRS 1.8 m telescope.

To explain the difference in $i-z$ colors of BHB stars and MSA stars, we examine high signal-to-noise spectra from the SDSS survey with particular attention paid to the 7000\AA-9200\AA \hspace{2 mm}range. We collect FITS images of all spectra in the SDSS survey which have a signal-to-noise ratio $\geq$ 50, are primary science readings, and classified as stars. Similar to \S 2, we select stars in the -0.3 $<$ $g-r$ $<$ 0.0 color range and define BHB stars as those in the 3.0 $<$ log($g$) $<$ 3.75 gravity range; MSA stars are defined as having 3.75 $<$ log($g$) $<$ 5.0. 

We construct two "super-spectra" by combining all spectra from each population. The spectra were individually shifted in wavelength such that the minimum of their Balmer-$\alpha$ absorption feature fell at 6563\AA \hspace{2 mm}to account for differing radial velocities. They were then normalized to their flux value at 7500\AA. It became apparent that, since our sample consisted of the whole -0.3 $<$ $g-r$ $<$ 0.0 color range (which spans from about 7500K to 10000K) a systematic difference in temperature needed to be accounted for (as our BHB sample is centered on this color range, but our MSA sample trends toward cooler, redder colors). We select BHB stars in the color range -0.15 $<$ $g-r$ $<$ -0.10 and MSA stars in the color range -0.17 $<$ $g-r$ $<$ -0.12-- these color ranges produced composite spectra which had approximately equal continua and Balmer line depths. We then bin the spectra into 2\AA \hspace{2 mm} wide bins and accept the median values of these bins as the composite value.

Figure 8 shows the well known Balmer-$\alpha$ absorption feature for our two spectra. We easily discern the differing profile widths and shapes that are the basis for differentiation of these types of stars in spectroscopic studies (e.g. Pier 1983, Clewley 2002; the BHB features are significantly slimmer than their MSA analogs). In the second frame we expand the Paschen features of these two types of stars. It is unsurprising that we see a similar effect in the $z$ band features-- the BHB features are significantly slimmer. Near a hydrogen absorption series limit, the features will overlap to form a quasi-continuum, thus the more intense Stark pressure broadening of the MSA stars makes their features wider and forces their pseudo-continua lower. This combination of slimmer absorption features and a higher Paschen continuum explains the difference in $z$ band magnitude for BHB and MSA stars at the same temperature.

An interesting corrollary to this result is the possibility of using only the Paschen features to separate BHB and MSA stars spectroscopically. To examine this, we consider only the area of the spectra between 8500\AA and 9000\AA; we only use this portion of the spectra since fitting to the entire spectra provides especially poor fits in the Paschen region. The continuum is defined to be the portion of the spectra exactly in between any two Paschen minima $\pm$5\AA \hspace{2 mm} (2 standard deviation outliers are discarded)-- the continuum is then fit to a tertiatic, which is divided out. The resulting normalized spectra are analyzed using an IRAF\footnote{IRAF is distributed by the National Optical Astronomy Observatories, which are operated by the Association of Universities for Research in Astronomy, Inc., under cooperative agreement with the National Science Foundation.} \citep{tod86} fitting routine. Included in Figure 8 is a plot of intensity vs the Voigt profile equivalent width of the Paschen line at $\sim$8596\AA \hspace{2 mm} (the Paschen 14 line) for the entire high signal to noise dataset. We see good separation of BHB and MSA stars in this plot. Notably, in this region, we see the BHB stars as having larger equivalent widths, which is counterintuitive. This is an effect of the pseudo-continuum shrinking the MSA features artificially owing to their broad and extended wings \citep{fre96}-- an effect corroborated by the correspondingly higher core intensities of the BHB stars.

%
%
\section{Conclusion}
In \S 3 we find, in agreement with \citet{bel10} and \citet{sir04}, that $u$ based color cuts may photometrically select out samples of BHB stars which are $\sim$74\% pure and $\sim$57\% complete. We also show that the unexplored $z$ band is capable of selecting out samples of BHB stars with similar accuracies (for structural studies, purity is more valuable than completeness), having achieved an estimated purity of $\sim$77\% based on spectroscopic data and globular cluster analysis. However, since the A-star separation by gravity is weaker in the $z$ band than in the $u$ band (as noted by Lenz et al. 1998), we obtained a lower completeness of $\sim$51\%. For increased accuracy, we suggest a combination of the two techniques, since they extract overlapping but non-identical sets.

In \S 4 we show that this color cut allows structure mapping in the Milky Way to large distances. Structures such as the northern Galactic portion of the Sagittarius tidal stream, which are on the very edge of F-turnoff detectability, are easily discerned in their radial completeness. Thus this method may be used for large-scale structural analysis in addition to high efficiency spectroscopic targeting. In \S 5 we show the spectroscopic reasoning for the $z$ band magnitude differences in BHB and MSA stars-- namely that Stark pressure broadening causes the continua of these stars to exist at fundamentally different levels. In short, we suggest that this algorithm is an excellent addition to existing photometric selection methods for structure characterization via BHB stars.

%
%
\section{Aknowledgements}
We thank C. J. Hansen, A. Koch and A. Pasquali for useful discussion regarding spectral analysis techniques and theory, and the referee for their helpful comments.

This work was supported by the Marie Curie Initial Training Networks grant number PITN-GA-2010-264895 ITN "Gaia Research for European Astronomy Training" and by Sonderforschungsbereich SFB 881 "The Milky Way System" (subproject A2 and A3)
of the German Research Foundation (DFG).

The SDSS-III Collaboration (www.sdss3.org) includes many institutions from around the globe. Funding for SDSS-III has been provided by the Alfred P. Sloan Foundation, the Participating Institutions, the National Science Foundation, and the U.S. Department of Energy. The SDSS-III is managed by the Astrophysical Research Consortium for the Participating Institutions of the SDSS-III Collaboration including the University of Arizona, the Brazilian Participation Group, Brookhaven National Laboratory, University of Cambridge, University of Florida, the French Participation Group, the German Participation Group, the Instituto de Astrofisica de Canarias, the Michigan State/Notre Dame/JINA Participation Group, Johns Hopkins University, Lawrence Berkeley National Laboratory, Max Planck Institute for Astrophysics, New Mexico State University, New York University, the Ohio State University, the Penn State University, University of Portsmouth, Princeton University,the Spanish Participation Group, University of Tokyo, the University of Utah,Vanderbilt University, University of Virginia, University of Washington, and YaleUniversity.

%
%

%
%

\begin{figure}
\epsscale{1}
\plotone{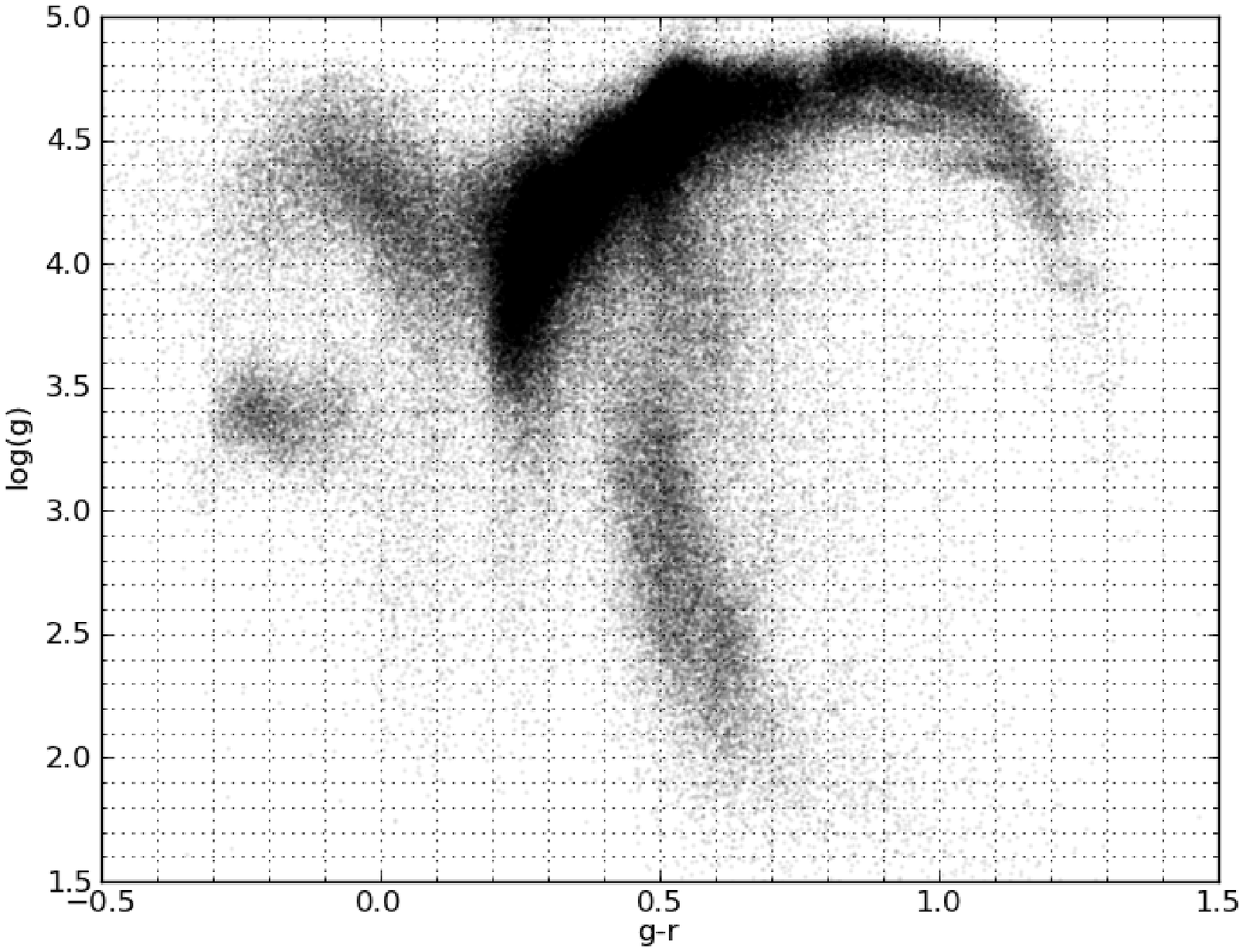}
\caption{Dereddened $g-r$ color versus log(g) as calculated by the SSPP pipeline for all spectroscopic sources with valid log(g) measurements. In the A-star colors (ranging from about $-0.3 <$ $g-r$ $<0.0$) we see two distinct stellar species: the lower surface gravity blue horizontal branch stars and the high surface gravity MSA stars. It is important to note that white dwarfs are largely absent in this diagram (as are quasars) since the SSPP adopted log(g) algorithm fails for these objects. \citet{lee08} estimate the average uncertainty in the log(g) measurement from the SSPP to be 0.29 dex.}\label{fig1}
\end{figure}

\begin{figure}
\epsscale{1}
\plotone{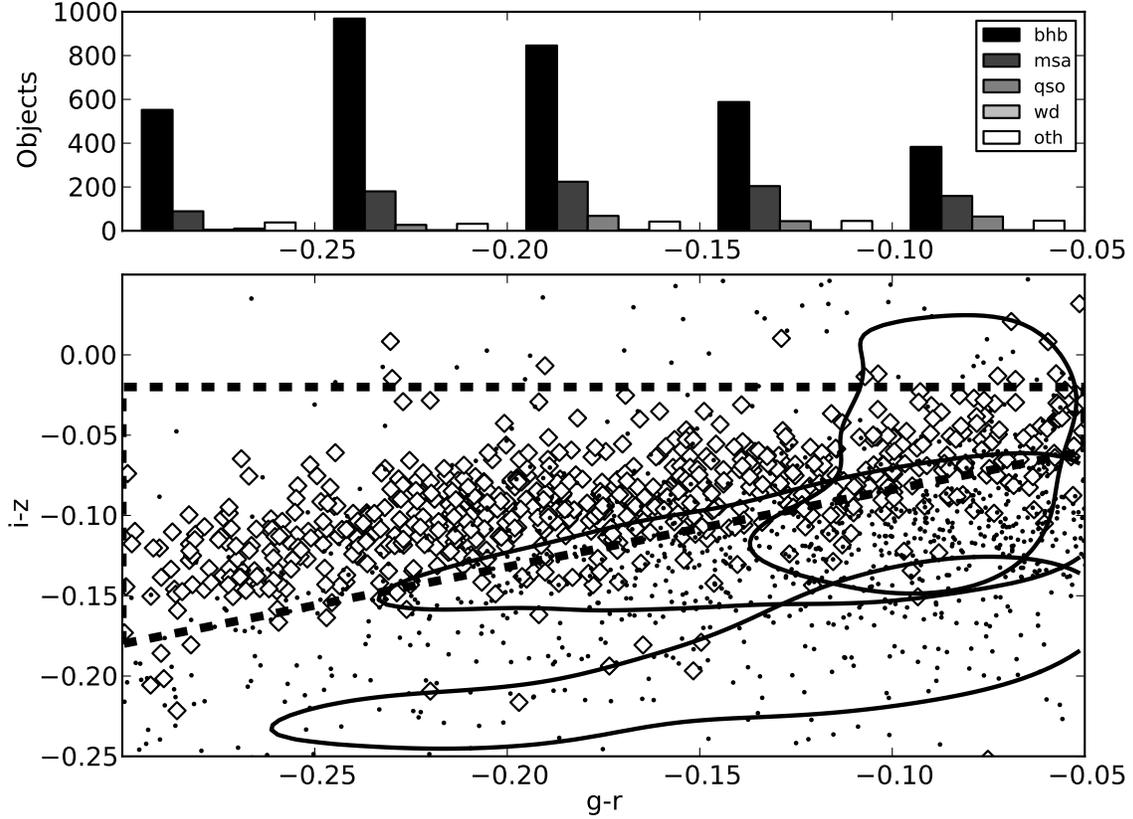}
\caption{The lower panel of this plot shows the distribution of objects in the $g-r$ versus $i-z$ color space. Plotted with diamonds are the BHBs (as determined by color and surface gravity) and contaminants to this population are plotted as points-- only 1 in 5 points from the entire dataset are plotted to avoid overcrowding. The contour lines show the general locations, in this color box, of quasars (qso, the reddest in $i-z$ with a locus of about $i-z$ = -0.07), main sequence A stars (msa, having a locus at around $i-z$ = -0.12) and white dwarfs (wd, the bluest, residing around $i-z$=-0.20): other contaminants (oth) are mostly A and F stars for which the SSPP failed to assign a gravity or assigned a gravity outside of the selection for either BHB or MSA described in \S 2. In the upper panel, the number and type of objects in the BHB selection box (the dashed polygon) are histogrammed as a function of $g-r$ color-- all of the dataset is presented in this frame. We see that hardly any white dwarfs pass this color cut.}\label{fig2} 
\end{figure}

\begin{figure}
\epsscale{1}
\plotone{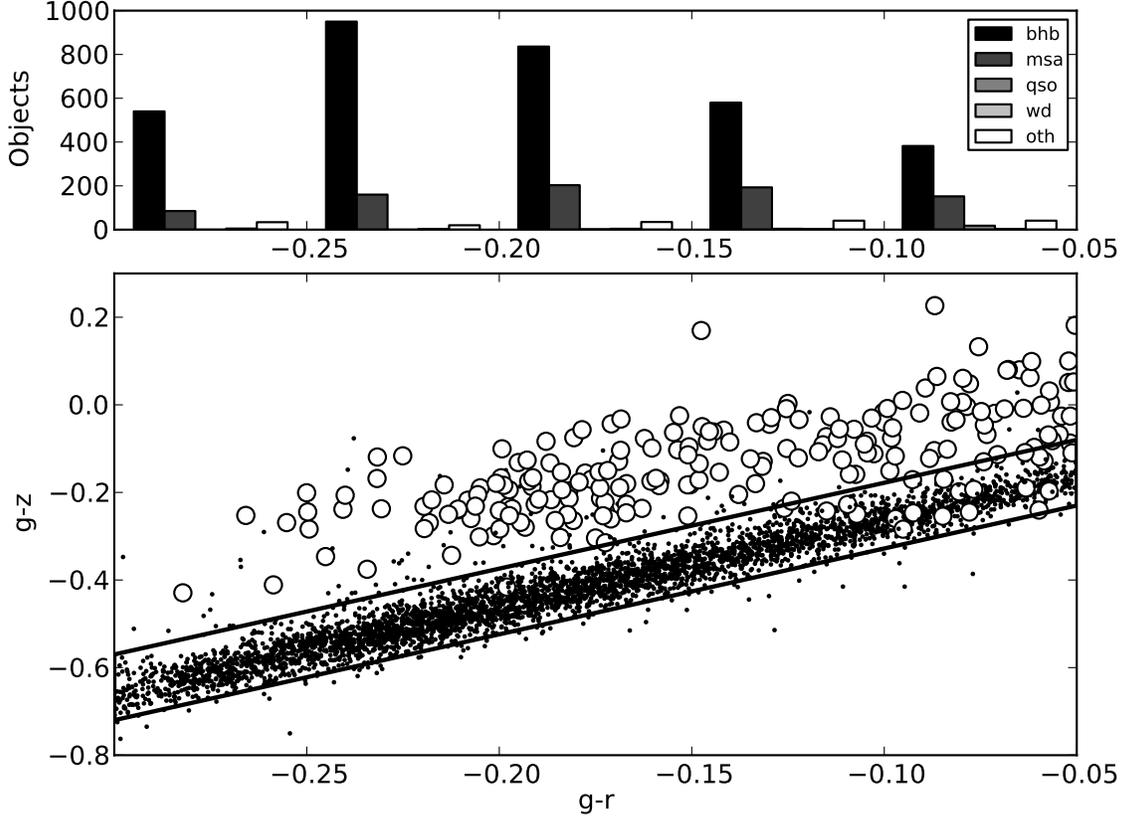}
\caption{The lower panel shows the distribution of stellar objects (BHBs, MSAs and white dwarfs; plotted as dots) in $g-r$ vs $g-z$ color space as compared to that of quasars (plotted as circles). Only data which pass the color-color cut shown in Figure 2 are presented here. The upper panel shows the number of objects in the BHB selection after this second color-color cut as a function of $g-r$ color. Comparison to Figure 2 shows that this second color cut is instrumental in removing quasars from the selection sample.}\label{fig3}
\end{figure}

\begin{figure}
\epsscale{1}
\plotone{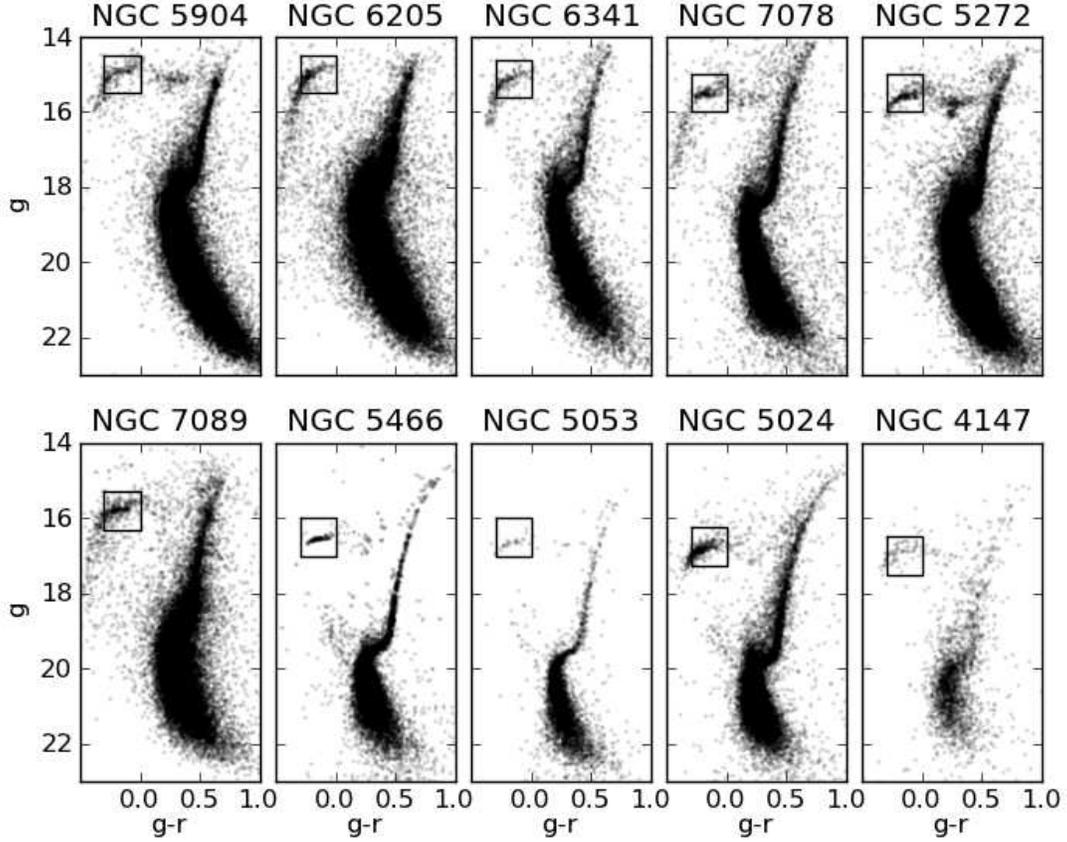}
\caption{Color-magnitude diagrams of 10 globular clusters found in the SDSS footprint, arranged by radial distance from the sun (Harris (1996) Catalog, 2010 revision values; Table 1). The drawn boxes indicate the general location (as selected by eye) of the "true" BHBs used for the completeness test described in \S 3. Each box extends 1 magnitude in $g$ from $g-r$ $=-0.3$ to $0.0$. Clusters such as NGC 7078 and NGC 5272 show RR Lyrae contamination on the red end of the box and clusters such as NGC 6205 and NGC 6341 show sub-dwarf contamination on its blue end.}\label{fig4}
\end{figure}

\begin{figure}
\epsscale{1}
\plotone{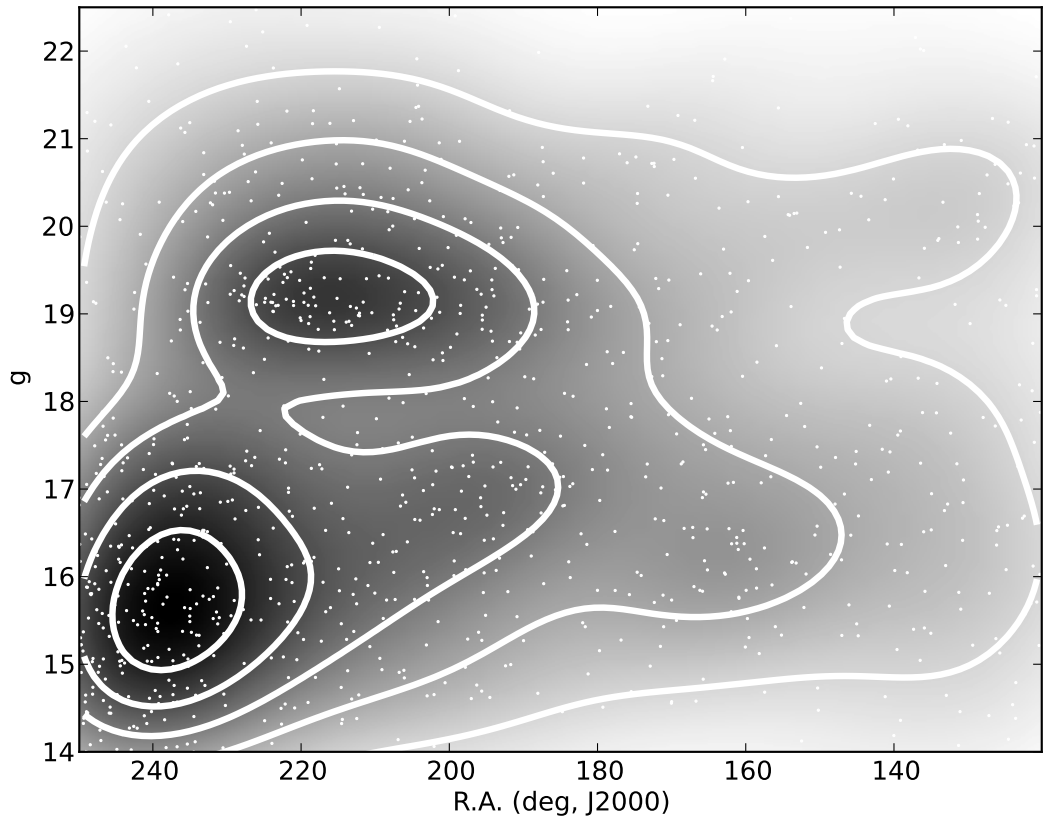}
\caption{The northern Galactic hemisphere section of the celestial equator (stripe 10). Photometric objects passing the color cuts shown in Figures 2 and 3 are presumed to be BHBs and are plotted here. The density of the points are shown via a gray-scaled density estimate which is accentuated by the white contours. We see the Hercules-Aquila cloud at R.A. $\approx$ 240$^{\circ}$ and $g$ $\approx$ 16; from R.A. $\approx$ 230$^{\circ}$ to $\approx$ 200$^{\circ}$ the Sagittarius dwarf spheroidal's tidal stream is apparent at $g$ $\approx$ 19; the Virgo overdensity can be seen at R.A. $\approx$ 200$^{\circ}$ and $g$ $\approx$ 17, the overdensity at R.A. $\approx$ 160$^{\circ}$ and $g$ $\approx$ 16 is noted in F-turnoff stars by \citet{new02}.}\label{fig5}
\end{figure}

\begin{figure}
\epsscale{1}
\plotone{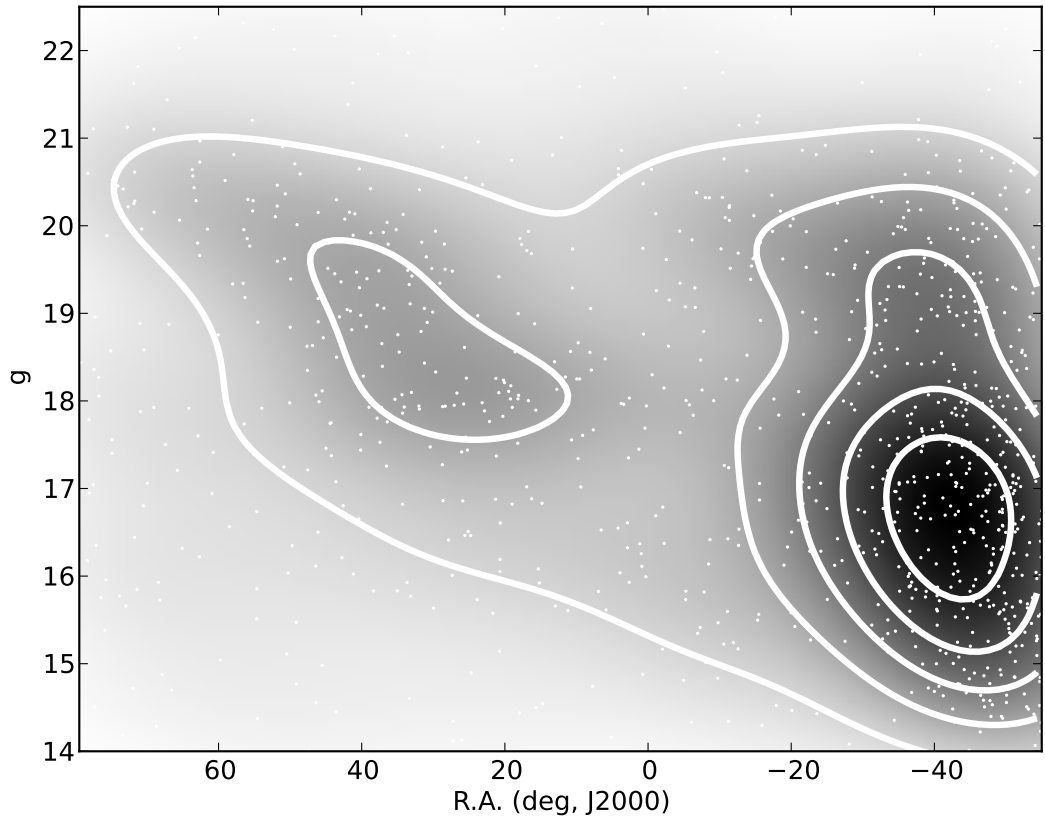}
\caption{The southern Galactic hemisphere section of the celestial equator (stripe 82)-- constructed in a similar fashion to Figure 5. Once again we see evidence for major known structures such as Sagittarius at R.A. $\approx$ 30$^{\circ}$ and $g$ $\approx$ 18 and the Hercules Aquila cloud at R.A $\approx$ -40$^{\circ}$ and $g$ $\approx$ 17.}\label{fig6}
\end{figure}

\begin{figure}
\epsscale{1}
\plotone{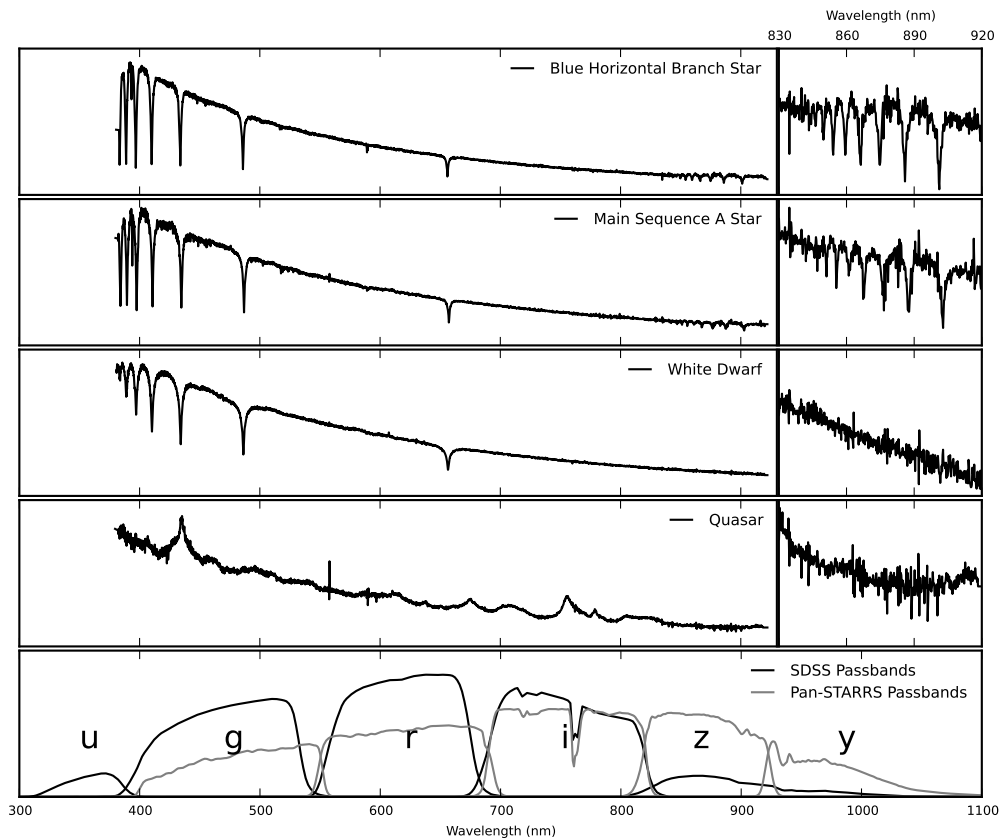}
\caption{The top four frames of this image are SDSS spectra of a typical BHB star, MSA star, white dwarf and low redshift quasar, respectively. The bottom frame depicts the SDSS passband functions ($ugriz$, from left to right) as described by \citet{doi10} and the Pan-STARRS passband functions ($grizy$, from left to right) provided on the Pan-STARRS website (http://svn.pan-starrs.ifa.hawaii.edu/trac/ipp/wiki/PS1\_Photometric\_System). In the spectra of the BHB and MSA stars, we see the Paschen features in the 830 nm to 920 nm range (which falls in the z-band of both systems; expanded to the right)-- these features will cause a difference in $z$ band based colors for BHB and MSA stars since they are susceptible to pressure broadening.}\label{fig7}
\end{figure}

\begin{figure}
\epsscale{1}
\plotone{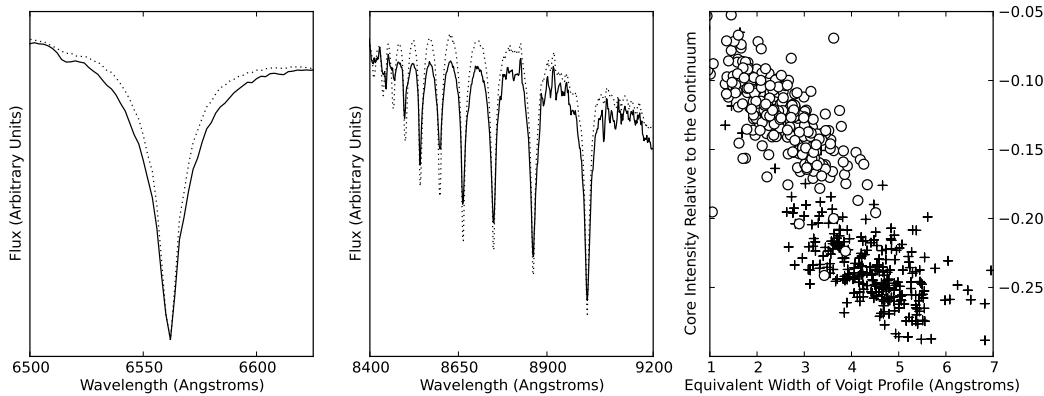}
\caption{The left frame shows the Balmer-$\alpha$ line for a composite MSA star (solid line) and a composite BHB star (dotted line). Due to Stark pressure broadening, these two absorption line profiles have different line widths and shapes which allows separation of these two species of star based on their differing surface gravities. The middle frame shows the Paschen absorption features in the $z$ band region. These features show a similar behavior with the BHB spectra (dotted line) having slimmer absorption lines than the MSA star (solid line). Since the continuum at this point is actually a quasi-continuum formed by the Paschen features flowing into each other, the BHB star appears to have a higher continuum than the MSA star (since broader features would push the effective continuum lower)-- this is the effect responsible for the differing $z$ band magnitudes. The rightmost frame depicts the difference in equivalent width versus core intensity for the Paschen feature residing at 8596\AA \hspace{2 mm}(the Paschen 14 line) for BHB stars (crosses; having 3.0 $<$ log($g$) $<$ 3.75) and MSA stars (circles; having 3.75 $<$ log($g$) $<$ 5.0). This plot was constructed using the high signal to noise spectra contributing the to combined spectra (described in \S 5). We see a distinct separation between these two species of stars in this parameter space; the pseudo-continuum is higher in the BHB stars which causes the feature to have a larger equivalent width and a higher core intensity.}\label{fig8}
\end{figure}

\clearpage

\begin{table}
\begin{center}
\caption{Globular Cluster Information.\label{tbl-1}}
\begin{tabular}{ccccccc}
\tableline\tableline
Cluster Name & R.A. & Dec & R$_\sun$ (kpc) & E(B$-$V) & R$_{tidal}$ (') & $g$ box \\
\tableline
NGC 4147 & 12 10 06.30 & +18 32 33.5 & 19.3 & 0.02 & 6.08 & 16.5-17.5 \\
NGC 5024 & 13 12 55.25 & +18 10 05.4 & 17.9 & 0.02 & 18.36 & 16.25-17.25 \\
NGC 5053 & 13 16 27.09 & +17 42 00.9 & 17.4 & 0.01 & 11.43 & 16-17 \\
NGC 5272 & 13 42 11.62 & +28 22 38.2 & 10.2 & 0.01 & 28.72 & 15-16 \\
NGC 5466 & 14 05 27.29 & +28 32 04.0 & 16.0 & 0.00 & 15.68 & 16-17 \\
NGC 5904 & 15 18 33.22 & +02 04 51.7 & 7.5 & 0.03 & 23.63 & 14.5-15.5 \\
NGC 6205 & 16 41 41.24 & +36 27 35.5 & 7.1 & 0.02 & 21.01 & 14.5-15.5 \\
NGC 6341 & 17 17 07.39 & +43 08 09.4 & 8.3 & 0.02 & 12.44 & 14.6-15.6 \\
NGC 7078 & 21 29 58.33 & +12 10 01.2 & 10.4 & 0.10 & 27.30 & 15-16 \\
NGC 7089 & 21 33 27.02 & -00 49 23.7 & 11.5 & 0.06 & 12.45 & 15.3-16.3 \\
\tableline
\end{tabular}
\tablecomments{All Information from the Harris (1996) Catalog, 2010 revision, excepting the g selection box. The $g$ selection box extends from $g-r$ = -0.3 to 0.0 in all cases.}
\end{center}
\end{table}

\clearpage
\begin{table}
\begin{center}
\caption{Globular Cluster Test Statistics.\label{tbl-2}}
\begin{tabular}{ccccccc}
\tableline\tableline
Cluster Name & [Fe/H] & Total BHBs & Purity$_{griz}$ & Completeness$_{griz}$ & Purity$_{ugr}$ & Completeness$_{ugr}$ \\
\tableline
NGC 5904 & -1.29 & 184 & 0.96 & 0.38 & 0.94 & 0.53 \\
NGC 5272 & -1.5 & 233 & 0.94 & 0.40 & 0.89 & 0.54 \\
NGC 6205 & -1.53 & 251 & 0.94 & 0.47 & 0.84 & 0.54 \\
NGC 7089 & -1.65 & 318 & 0.94 & 0.37 & 0.94 & 0.36 \\
NGC 4147 & -1.8 & 60 & 0.96 & 0.45 & 0.94 & 0.48 \\
NGC 5466 & -1.98 & 115 & 1.00 & 0.84 & 0.96 & 0.84 \\
NGC 5024 & -2.1 & 351 & 0.96 & 0.70 & 0.94 & 0.70 \\
NGC 5053 & -2.27 & 28 & 0.92 & 0.82 & 0.96 & 0.79 \\
NGC 6341 & -2.31 & 154 & 0.98 & 0.34 & 0.96 & 0.60 \\
NGC 7078 & -2.37 & 194 & 0.88 & 0.57 & 0.90 & 0.58 \\
Total & & 1888 & 0.95 & 0.51 & 0.92 & 0.57 \\
\tableline
\tableline
\end{tabular}
\tablecomments{This table shows the accuracy and effectiveness of the two tests in extracting BHB stars from the chosen globular clusters. Metallicities from the Harris (1996) Catalog, 2010 revision, are included and the globular clusters are sorted by this property-- one can see a slight trend in both tests to more effectively select BHB stars at lower metallicities, especially in the area around -2.0 dex. This is most likely an effect of our color cut being formulated by the color distribution of halo BHB stars spectroscopically observed by the SDSS, which have an average metallicity of -2.0 dex \citep{xue08}.}
\end{center}
\end{table}


\begin{thebibliography}{}
\bibitem[Abazajian et al.(2009)]{aba09} Abazajian, K.~N., 
Adelman-McCarthy, J.~K., Ag{\"u}eros, M.~A., et al.\ 2009, \apjs, 182, 543 

\bibitem[Aihara et al.(2011)]{aih11} Aihara, H., Allende 
Prieto, C., An, D., et al.\ 2011, \apjs, 193, 29 

\bibitem[An et al.(2008)]{an08} An, D., Johnson, J.~A., 
Clem, J.~L., et al.\ 2008, \apjs, 179, 326 

\bibitem[Bell et al.(2010)]{bel10} Bell, E.~F., Xue, X.~X., 
Rix, H.-W., Ruhland, C., \& Hogg, D.~W.\ 2010, \aj, 140, 1850 

\bibitem[Belokurov et al.(2007)]{bel07} Belokurov, V., Evans, 
N.~W., Bell, E.~F., et al.\ 2007, \apjl, 657, L89 

\bibitem[Clewley et al.(2002)]{cle02} Clewley, L., Warren, 
S.~J., Hewett, P.~C., et al.\ 2002, \mnras, 337, 87 

\bibitem[Clewley 
\& Jarvis(2006)]{cle06} Clewley, L., \& Jarvis, M.~J.\ 2006, \mnras, 368, 310 

\bibitem[Doi et al.(2010)]{doi10} Doi, M., Tanaka, M., 
Fukugita, M., et al.\ 2010, \aj, 139, 1628 

\bibitem[Dotter et al.(2010)]{dot10} Dotter, A., Sarajedini, 
A., Anderson, J., et al.\ 2010, \apj, 708, 698 

\bibitem[Fremat et al.(1996)]{fre96} Fremat, Y., Houziaux, 
L., \& Andrillat, Y.\ 1996, \mnras, 279, 25 

\bibitem[Frieman et al.(2008)]{fri08} Frieman, J.~A., 
Bassett, B., Becker, A., et al.\ 2008, \aj, 135, 338 

\bibitem[Fukugita et al.(1996)]{fuk96} Fukugita, M., 
Ichikawa, T., Gunn, J.~E., et al.\ 1996, \aj, 111, 1748 

\bibitem[Gunn et al.(1998)]{gun98} Gunn, J.~E., Carr, M., 
Rockosi, C., et al.\ 1998, \aj, 116, 3040 

\bibitem[Harris(1996)]{har96} Harris, W.~E.\ 1996, \aj, 112, 
1487 

\bibitem[Jordi 
\& Grebel(2010)]{jor10} Jordi, K., \& Grebel, E.~K.\ 2010, \aap, 522, A71 

\bibitem[Kaiser et al.(2010)]{kai10} Kaiser, N., Burgett, W., 
Chambers, K., et al.\ 2010, \procspie, 7733,  

\bibitem[Lee et al.(2008)]{lee08} Lee, Y.~S., Beers, T.~C., 
Sivarani, T., et al.\ 2008, \aj, 136, 2022 

\bibitem[Lenz et al.(1998)]{len98} Lenz, D.~D., Newberg, J., 
Rosner, R., Richards, G.~T., \& Stoughton, C.\ 1998, \apjs, 119, 121 

\bibitem[Moultaka et al.(2004)]{mou04} Moultaka, J., 
Ilovaisky, S.~A., Prugniel, P., \& Soubiran, C.\ 2004, \pasp, 116, 693

\bibitem[Newberg et al.(2002)]{new02} Newberg, H.~J., Yanny, 
B., Rockosi, C., et al.\ 2002, \apj, 569, 245 

\bibitem[Pier(1983)]{pie83} Pier, J.~R.\ 1983, \apjs, 53, 791 

\bibitem[Richards et al.(2009)]{2009ApJS..180...67R} Richards, G.~T., 
Myers, A.~D., Gray, A.~G., et al.\ 2009, \apjs, 180, 67 

\bibitem[Ruhland et al.(2011)]{ruh11} Ruhland, C., Bell, 
E.~F., Rix, H.-W., \& Xue, X.-X.\ 2011, \apj, 731, 119

\bibitem[Sesar et al.(2007)]{ses07} Sesar, B., Ivezi{\'c}, 
{\v Z}., Lupton, R.~H., et al.\ 2007, \aj, 134, 2236 

\bibitem[Schlegel et al.(1998)]{sch98} Schlegel, D.~J., 
Finkbeiner, D.~P., \& Davis, M.\ 1998, \apj, 500, 525 

\bibitem[Sirko et al.(2004)]{sir04} Sirko, E., Goodman, J., 
Knapp, G.~R., et al.\ 2004, \aj, 127, 899 

\bibitem[Stoughton et al.(2002)]{sto02} Stoughton, C., 
Lupton, R.~H., Bernardi, M., et al.\ 2002, \aj, 123, 485 

\bibitem[Stubbs et al.(2010)]{stu10} Stubbs, C.~W., Doherty, 
P., Cramer, C., et al.\ 2010, \apjs, 191, 376 

\bibitem[Tody(1986)]{tod86} Tody, D.\ 1986, \procspie, 627, 
733 

\bibitem[Vivas et al.(2001)]{viv01} Vivas, A.~K., Zinn, R., 
Andrews, P., et al.\ 2001, \apjl, 554, L33 

\bibitem[Wilhelm et al.(1999)]{wil99} Wilhelm, R., Beers, 
T.~C., Sommer-Larsen, J., et al.\ 1999, \aj, 117, 2329 

\bibitem[Xue et al.(2008)]{xue08} Xue, X.~X., Rix, H.~W., 
Zhao, G., et al.\ 2008, \apj, 684, 1143 

\bibitem[Yanny et al.(2000)]{yan00} Yanny, B., Newberg, 
H.~J., Kent, S., et al.\ 2000, \apj, 540, 825 

\bibitem[Yanny et al.(2009)]{yan09} Yanny, B., Rockosi, C., 
Newberg, H.~J., et al.\ 2009, \aj, 137, 4377 

\bibitem[York et al.(2000)]{yor00} York, D.~G., Adelman, J., 
Anderson, J.~E., Jr., et al.\ 2000, \aj, 120, 1579 

\end{thebibliography}
\end{document}